%% file: main.tex
\documentclass[10pt,journal,compsoc]{IEEEtran}

\usepackage{amssymb}
\usepackage{pdfpages}
\usepackage{enumitem}
\usepackage{bbm}
\usepackage{multirow}
\usepackage{algorithm,algpseudocode}
\usepackage{url}
\usepackage{gensymb}
\input{./TeX/commands}

\usepackage{array}
\newcommand{\PreserveBackslash}[1]{\let\temp=\\#1\let\\=\temp}
\newcolumntype{C}[1]{>{\PreserveBackslash\centering}p{#1}}
\newcolumntype{R}[1]{>{\PreserveBackslash\raggedleft}p{#1}}
\newcolumntype{L}[1]{>{\PreserveBackslash\raggedright}p{#1}}

\begin{document}

\title{Dynamic Low-light Imaging with \\ Quanta Image Sensors}
\author{Yiheng~Chi, Abhiram Gnanasambandam, Vladlen~Koltun, and Stanley H. Chan
\thanks{Y. Chi, A. Gnanasambandam and S. H. Chan are with the School of Electrical and Computer Engineering, Purdue University, West Lafayette, IN 47907, USA. Email: \texttt{\{agnanasa, chi14, stanchan\}@purdue.edu}. V. Koltun is with Intel Labs, Santa Clara, CA 95054, USA.

This work is supported, in part, by the National Science Foundation under grant CCF-1718007.

This paper is presented in the 16-th European Conference on Computer Vision (ECCV), Glasgow, United Kingdom, August 2020.}
}
\graphicspath{{./pix/}}

\IEEEtitleabstractindextext{\begin{abstract}
Imaging in low light is difficult because the number of photons arriving at the sensor is low. Imaging dynamic scenes in low-light environments is even more difficult because as the scene moves, pixels in adjacent frames need to be aligned before they can be denoised. Conventional CMOS image sensors (CIS) are at a particular disadvantage in dynamic low-light settings because the exposure cannot be too short lest the read noise overwhelms the signal. We propose a solution using Quanta Image Sensors (QIS) and present a new image reconstruction algorithm. QIS are single-photon image sensors with photon counting capabilities. Studies over the past decade have confirmed the effectiveness of QIS for low-light imaging but reconstruction algorithms for dynamic scenes in low light remain an open problem. We fill the gap by proposing a student-teacher training protocol that transfers knowledge from a motion teacher and a denoising teacher to a student network. We show that dynamic scenes can be reconstructed from a burst of frames at a photon level of 1 photon per pixel per frame. Experimental results confirm the advantages of the proposed method compared to existing methods.
\end{abstract}

}

\maketitle
\section{Introduction}
Imaging in photon-starved situations is one of the biggest technological challenges for applications such as security, robotics, autonomous cars, and health care. However, the growing demand for higher resolution, smaller pixels, and smaller form factors have limited the photon sensing area of the sensors. This, in turn, puts a fundamental limit on the signal-to-noise ratio that the sensors can achieve. Over the past few years, there is an increasing amount of effort in developing alternative sensors that have photon-counting ability. Quanta Image Sensors (QIS) are one of these new types of image sensors that can count individual photons at a very high frame rate and have a high spatial resolution \cite{ma2015pump,ma2017photon}. Various prototype QIS have been reported, and numerous studies have confirmed their capability for high speed imaging \cite{burri2014architecture}, high dynamic range imaging \cite{gnanasambandamhigh,elgendy2017optimal}, color imaging \cite{Gnanasambandam:19,elgendy2019color}, and tracking \cite{gyongy2018single}.

Despite the increasing literature on QIS sensor development \cite{fossum2013modeling,ma2015pump,ma2017photon} and signal processing algorithms \cite{yang2010optimal,chandramouli2019bit}, one of the most difficult problems in QIS is image reconstruction for \emph{dynamic} scenes. Image reconstruction for dynamic scenes is important for broad adoption of QIS: solving the problem can open the door to a wide range of low-light applications such as videography, moving object detection, non-stationary facial recognition, etc. However, motion in low light is difficult because it must deal with two types of distortions: low light causes shot noise which is random and affects the entire image, whereas motion causes geometric warping which is often local. In this paper, we address this problem with a new algorithm.

\begin{figure*}[t]
    \centering
    \begin{tabular}{ccc}
    \hspace{-2ex}\includegraphics[width=0.33\linewidth]{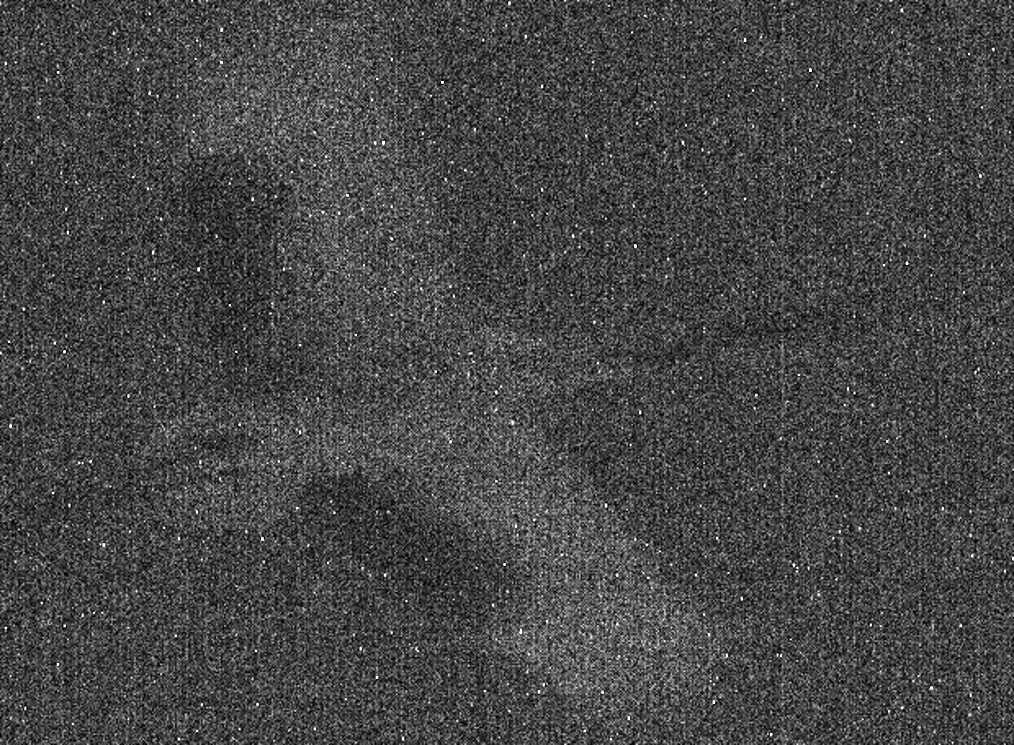} &
    \hspace{-2ex}\includegraphics[width=0.33\linewidth]{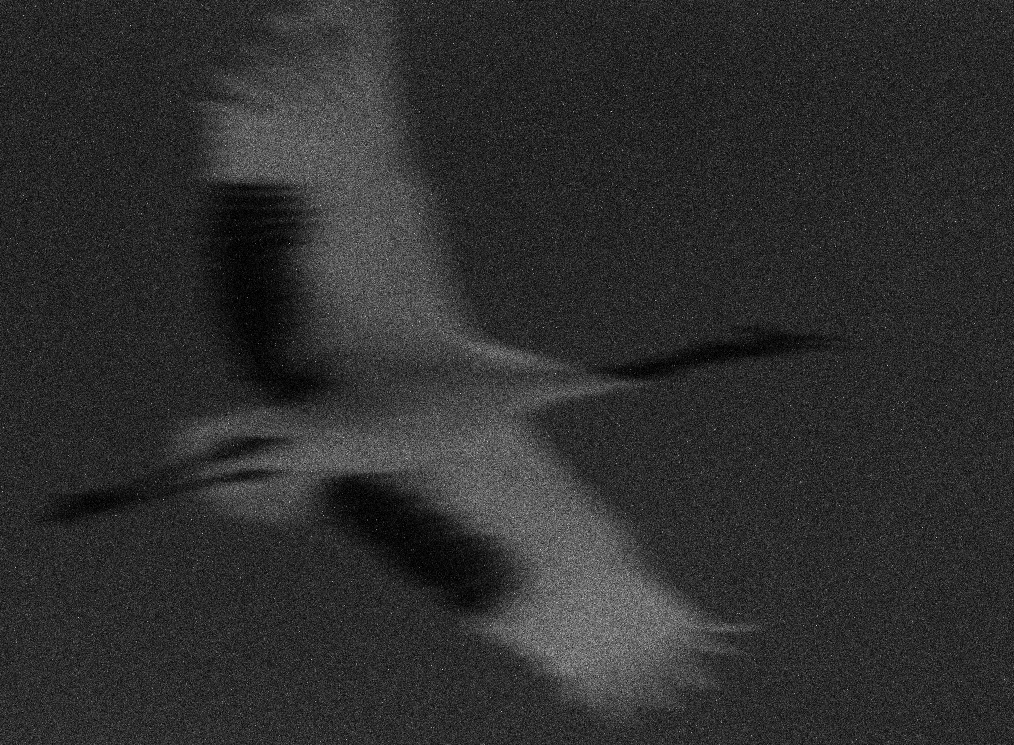} &
    \hspace{-2ex}\includegraphics[width=0.33\linewidth]{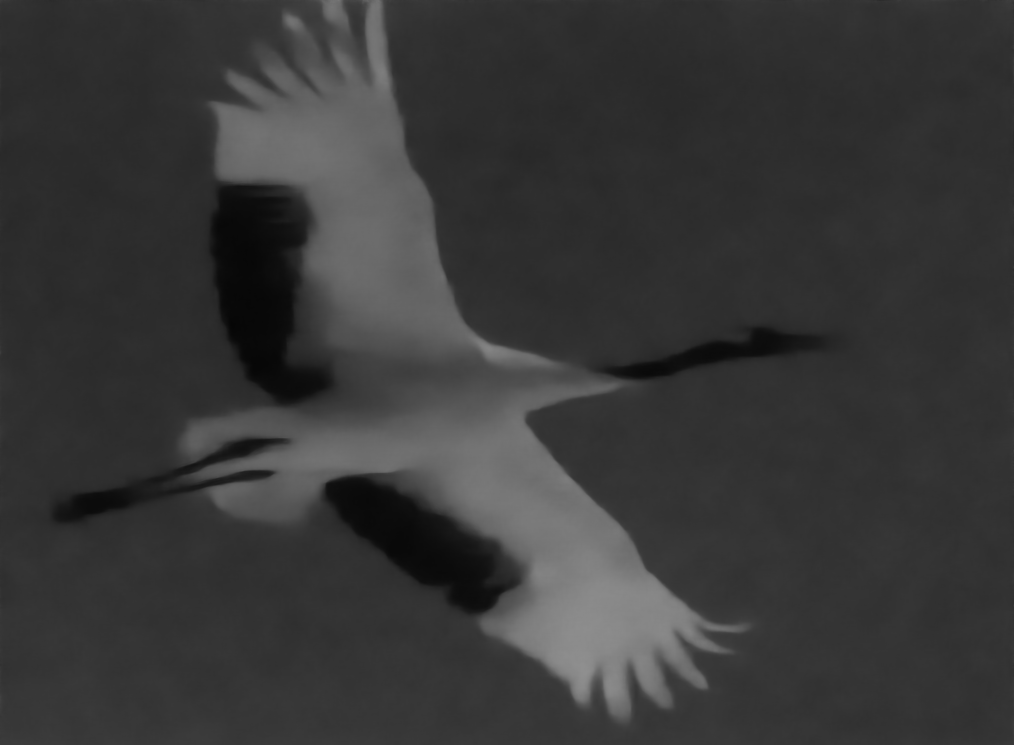} \\
    \small{(a) Real image by CIS} & \small{(b) Real image by QIS} & \small{(c) Our reconstruction} \\
    \small{avg of 8 frames, 0.5 ppp} & \small{avg of 8 frames, 0.5 ppp} & \small{using 8 QIS frames}
    \end{tabular}
    \caption{\textbf{Goal of this paper}. The images above are the \emph{real} captures by a CMOS Image Sensor (CIS) and a QIS prototype \cite{Gnanasambandam:19} at the same photon level of 0.5 photons per pixel (ppp) per frame. The strong shot noise and read noise of CIS makes signal acquisition difficult, whereas the QIS can obtain a better image. Using the proposed method, we are able to reconstruct images with dynamic content.}
    \label{fig:CIS_vs_QIS}
\end{figure*}

\begin{figure*}[t]
\centering
    \begin{tabular}{cccc}
    \hspace{-2ex}\includegraphics[width=0.24\linewidth]{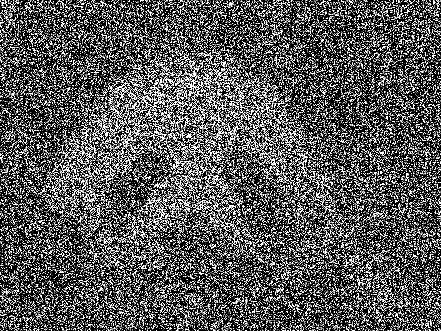}&
    \hspace{-2ex}\includegraphics[width=0.24\linewidth]{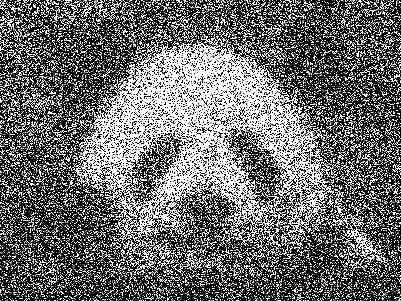}&
    \hspace{-2ex}\includegraphics[width=0.24\linewidth]{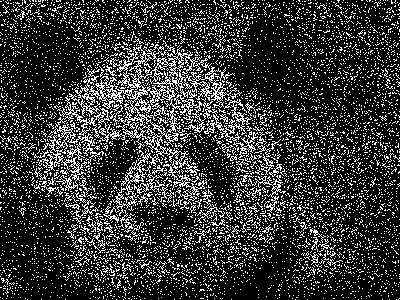}&
    \hspace{-2ex}\includegraphics[width=0.24\linewidth]{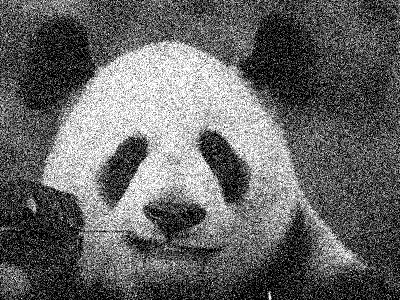}\\
(a) CIS (real) & (b) QIS (real) &  (c) Ideal sensor &  (d) i.i.d. Gaussian \\
   0.25 ppp    & 0.25 ppp & 0.25 ppp & $\sigma = 50/255$
    \end{tabular}
    \caption{\textbf{Photon level and sensor limitations.} (a) and (b) show a pair of real images captured by CIS and QIS at 0.25 ppp. (c) shows a simulated image acquired by an ``ideal sensor'' which is free of read noise and dark current. The random shot noise in this ideal image suggests that although QIS has higher sensitivity than CIS, image reconstruction algorithms still play a critical role because there is a fundamental limit due to the Poisson statistics. (d) shows an image distorted by i.i.d. Gaussian noise of a strength $\sigma = 50/255$, somewhat high in the denoising literature.}
\label{fig:Noisy}
\end{figure*}

\fref{fig:CIS_vs_QIS} summarizes our objective. \fref{fig:CIS_vs_QIS}(a) shows real data captured by a conventional CMOS image sensor (CIS). The photon level is 0.5 photons per pixel (ppp). \fref{fig:CIS_vs_QIS}(b) shows the data captured by a QIS at the same photon level. To illustrate the effect of motion, we show the average of 8 consecutive frames. \fref{fig:CIS_vs_QIS}(c) shows the result of the proposed image reconstruction algorithm applied to the 8 QIS frames. Although the scene is in motion, the presented approach recovers most of the image details. This brings out the two contributions of this paper:

\begin{enumerate}
\item[(i)] We demonstrate low-light image reconstruction of dynamic scenes at a photon level of 1 photon per pixel (ppp) per frame. This is lower than most of the results reported in the computational photography literature.
\item[(ii)] We propose a student-teacher framework and show that this training method is effective in handling noise and motion simultaneously.
\end{enumerate}

\section{Background}
\label{sec: Background}
\subsection{Quanta Image Sensors}
Quanta Image Sensors (QIS) were originally proposed in 2005 as a candidate solution for the shrinking pixel problem \cite{fossum200611,fossum2005gigapixel}. The idea is to partition a CIS pixel into many tiny cells called ``jots'' where each jot is a single-photon detector. By oversampling the scene in space and time, the underlying image can be recovered using a carefully designed image reconstruction algorithm. Numerous studies have analyzed the theoretical properties of these sensors, including their performance limit \cite{yang2011bits}, photon statistics \cite{fossum2013modeling}, threshold analysis \cite{elgendy2017optimal}, dynamic range \cite{gnanasambandamhigh}, and color filter array \cite{elgendy2019color}. On the hardware side, a number of prototypes have become available \cite{dutton2015spad,dutton2014320,ma2015pump}. The prototype QIS we use in this paper is based on \cite{ma2017photon}.

As photon counting devices, QIS share many similarities with single-photon avalanche diodes (SPAD) \cite{dutton2015spad}. However, SPAD amplify signals using avalanche multiplication. This requires a high electrical voltage (typically higher than 20V) to accelerate the photoelectron. Because avalanche multiplication requires space for electrons to multiply, SPAD have high dark current ($>10e^-/\text{pix}/\text{s}$), large pitch ($>5\mu$m), low fill-factor ($<70\%$), and low quantum efficiency ($<50\%$). In contrast, QIS do not require avalanche multiplication. They have significantly better fill-factor, quantum efficiency, dark current, and read noise. SPAD are excellent candidates for resolving time-stamps, e.g., time-of-flight applications \cite{Gupta_ICCV2019,o2017reconstructing,lindell2018single,callenberg2019emccd,gariepy2015single}, although new studies have shown other applications \cite{Ma_SIGGRAPH20}. QIS have higher resolution which makes them suitable for low-light photography. Recent literature provides a more detailed comparison \cite{Gnanasambandam:19}.

\subsection{How Dark is One Photon Per Pixel?}
All photon levels in this paper are measured in terms of photons per pixel (ppp). ``Photons per pixel'' is the average number of photons a pixel detects during the exposure period. We use photons per pixel as the metric because the amount of photons detected by a sensor depends on the exposure time and sensor size. A large sensor can collect more photons, and longer exposure time would allow more photons to arrive at the sensor. Therefore, even for the same scene with the same illuminance (measured in lux), the number of photons per pixel seen by two sensors can be different. To give readers an idea of the amount of noise we are dealing with in this paper, \fref{fig:Noisy}(a,b) shows a pair of real images captured by CIS and QIS at 0.25 ppp. Note that the signal at this photon level is significantly worse than what is commonly considered ``heavy noise'' in the denoising literature, illustrated in \fref{fig:Noisy}(d). We should also highlight that while QIS is a better sensor, at low light the signal-to-noise ratio is upper bounded by the fundamental limit of the Poisson process. As shown in \fref{fig:Noisy}(c), an ideal sensor with zero read noise and zero dark current will still produce an image contaminated by shot noise. Therefore, reconstruction algorithms are needed to recover the images even though QIS have higher photon sensitivity than CIS.

\subsection{Related Work}

\textbf{QIS Image Reconstruction}. Image reconstruction for QIS is challenging because of the unique Poisson-Gaussian statistics of the sensor. Early reconstruction techniques are based on solving maximum-likelihoods using gradient descent \cite{yang2010optimal}, dynamic programming \cite{yang2009image}, and convex optimization techniques \cite{chan2014efficient,chan2016plug}. The first non-iterative algorithm for QIS image reconstruction was proposed by Chan et al. \cite{Chan16}. It was shown that if one assumes spatial independence, then the truncated Poisson likelihood can be simplified to Binomial. Consequently, the Anscombe binomial transform can be used to stabilize the variance, and off-the-shelf denoising (e.g., BM3D \cite{Dabov_Foi_Katkovnik_2007}) can be used to denoise the image. Choi et al. \cite{choi2018image} followed the idea by replacing the denoiser with a deep neural network. Alternative solutions using end-to-end deep neural networks have also been proposed for QIS \cite{remez2016picture} and SPAD \cite{chandramouli2019bit}. To the best of our knowledge, ours is the first dynamic scene reconstruction for QIS.

\begin{figure*}[t]
\centering
\begin{tabular}{cccc}
    \hspace{-2ex}\includegraphics[width=0.23\linewidth]{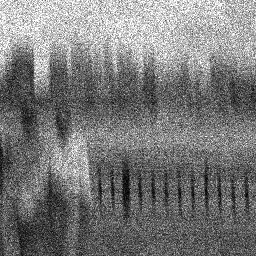} &
    \hspace{-2ex}\includegraphics[width=0.23\linewidth]{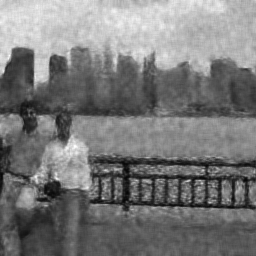} &
    \hspace{-2ex}\includegraphics[width=0.23\linewidth]{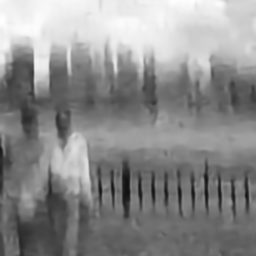} &
    \hspace{-2ex}\includegraphics[width=0.23\linewidth]{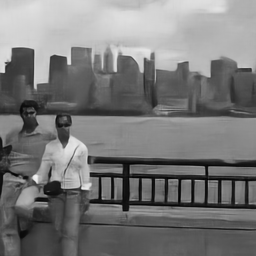} \\
    (a) QIS raw data & (b) KPN \cite{mildenhall2018burst} & (c) sRED \cite{mao2016image} & (d) Ours \\
    \small{8-frame avg} & 23.09 dB & 17.74 dB & 26.74 dB
\end{tabular}
\vspace{-2ex}
\caption{\textbf{The dilemma of noise and motion}. (a) A simulated QIS sequence at 2 ppp, averaged over 8 frames. (b) Result of Kernel Prediction Network (KPN) \cite{mildenhall2018burst}, a burst photography method that handles motion. (c) Result of a single-frame image denoiser sRED \cite{mao2016image} applied to the 8-frame avg. (d) Result of our proposed method.}
\label{fig:dilemma}
\end{figure*}

\textbf{Low-light Denoising}. The majority of existing denoising algorithms are designed for CIS. Single-frame image denoising methods are abundant, e.g., non-local means \cite{Buades_Coll_2005_Journal}, BM3D \cite{Dabov_Foi_Katkovnik_2007}, Poisson denoising \cite{makitalo2010optimal}, and many others \cite{malm2007adaptive,hu2014deblurring,guo2016lime, fu2018retinex}. On the deep neural network side, there are numerous networks dedicated to single-image denoising \cite{remez2017deep, zhang2017beyond,lore2017llnet,zhang2018ffdnet}. However, recent benchmark experiments found that BM3D is often better than deep learning methods for real sensor data \cite{plotz2017benchmarking, xu2018real}. Specific to low-light imaging, Chen et al. \cite{chen2018learning,chen2019seeing} observed that by modeling the entire image and signal processing pipeline using an end-to-end network, better reconstruction results can be obtained from the raw sensor data. However, since the images are still captured by CIS, the photon levels are much higher than what we study in this paper.

For dynamic scenes, extensions of the static methods to videos are available, e.g., based on non-local means \cite{davy2018non,sutour2014adaptive,buades2005denoising}, optical flow \cite{liu2010high,werlberger2011optical,liu2014fast}, and sparse representation \cite{protter2008image,ji2010robust}. The most relevant approach for this paper is the burst photography technique \cite{hasinoff2016burst}, which can be traced back to earlier methods based on optical flow \cite{liu2010high,liu2014fast,joshi2010seeing}. Recent reports on burst photography have focused on using deep neural networks \cite{xia2019basis,godard2018deep,kokkinos2019iterative,aittala2018burst}. Among these, the kernel prediction network (KPN) by Mildenhall et al.~\cite{mildenhall2018burst} is the most relevant work for us. However, as we will demonstrate later in the paper, the performance of KPN is not as satisfactory in the extreme noise conditions we deal with.

\section{Method}
\label{sec: method}
The proposed method consists of the QIS and a new image reconstruction algorithm. Before we discuss the algorithm, we first discuss how images are formed on QIS, as well as the challenges of imaging dynamic scenes in low-light. After that, we discuss the proposed solution using student-teacher learning, and the intuitions behind the method.

\subsection{QIS Imaging Model}
We now present the image formation model. Our model is based on the prototype QIS reported in \cite{ma2017photon} and is more detailed than the models used in existing literature such as \cite{Chan16,yang2011bits}.

As light travels from the scene to the sensor, the main mathematical model is the Poisson process which describes how photons arrive. However, due to various sources of distortions, the measured QIS signal, $\vx_\text{QIS}$, is given by

\begin{align}
\underset{\text{observed}}{ \underbrace{\vx_\text{\tiny{QIS}}}}
&= \text{ADC}\bigg\{ \underset{
\text{photon arrival}}{\underbrace{\text{Poisson}}}\bigg( \underset{\text{sensor gain}}{\underbrace{\alpha}} \;\cdot\; \big( \underset{\text{scene}}{ \underbrace{\vx_{\text{true}}} } + \ldots \notag \\
&\qquad\qquad +
\underset{\text{dark current}}{\underbrace{\veta_{\text{dc}}}} \big)\bigg) +
\underset{\text{read noise}}{\underbrace{\veta_{\text{r}}}}\bigg\}.
\label{eq: QIS equation}
\end{align}

Here we assume that the sensor is monochromatic because the real data reported in this paper are based on a monochromatic prototype QIS. To simulate color data we need to include a sub-sampling step to model the color filter array. $\veta_\text{dc}$ denotes the dark current and $\veta_{\text{r}}$ denotes the read noise arising from the read-out circuit. The analog-to-digital converter (ADC) describes the sensor output. In single-bit QIS, the output is a binary signal obtained by thresholding the Poisson count \cite{elgendy2017optimal}. In multi-bit QIS, the output is the Poisson count clipped to the maximum number of bits. To image a dynamic scene, we use QIS to collect a stack of short-exposure frames. Akin to previous work~\cite{Chan16,yang2011bits}, we assume that noise is independent over time.

For the prototype sensor we use in this paper, the dark current $\veta_{\text{dc}}$ in Equation \eref{eq: QIS equation} has an average value of 0.0068$e^-$/pix/s and the read noise $\veta_{\text{r}}$  takes the value of $0.25e^{-}$/pix \cite{ma2017photon}. The sensor gain $\alpha$ controls the exposure time and the dynamic range, which changes from scene to scene. For all experiments we conduct in this paper, the analog-to-digital conversion is 3-bit. The spatial resolution of the sensor is $1024\times 1024$, although we typically crop regions of the image for analysis.

\subsection{The Dilemma of Noise and Motion}
At the heart of dynamic image reconstruction is the coexistence of noise and motion. The dilemma here is that they are intertwined. To remove noise in a dynamic scene, we often need to either align the frames or construct a steerable kernel over the space-time volume. The alignment step is roughly equivalent to estimating optical flow \cite{horn1981determining}, whereas constructing the steerable kernel is equivalent to non-local means \cite{sutour2014adaptive,buades2005denoising} or kernel prediction \cite{mildenhall2018burst}. However, if the images are contaminated by noise, then both optical flow and kernel prediction will fail. When this step fails, denoising will be difficult because we will not be able to easily find neighboring patches for filtering.

Existing algorithms in the denoising literature can usually only handle one of the two situations. For example, the kernel prediction network (KPN) \cite{mildenhall2018burst} can extract motion information from a dynamic scene but its performance drops when noise becomes heavy. Similarly, the residual encoder-decoder networks REDNet \cite{mao2016image} and DnCNN \cite{zhang2017beyond} are designed for static scenes. In \fref{fig:dilemma}, we show the results of a synthetic experiment. The results illustrate the limitations of the motion-based KPN \cite{mildenhall2018burst} and the single-frame REDNet (sRED) \cite{mao2016image}. Our goal is to leverage the strengths of both.

\begin{figure*}[t]
    \centering
    \includegraphics[width=\linewidth]{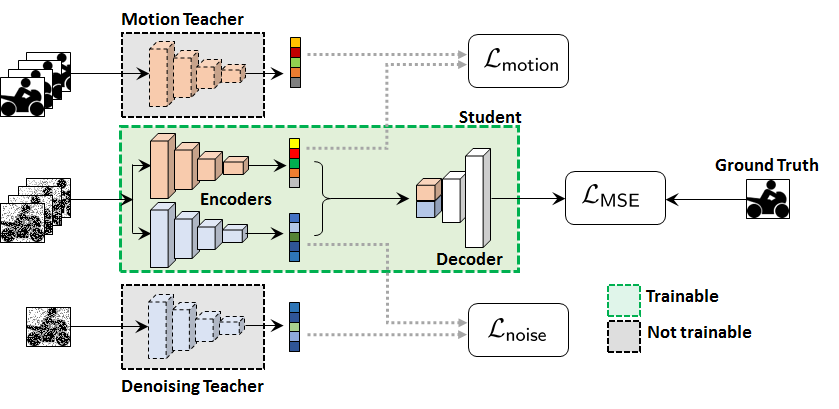}
    \vspace{-6ex}
    \caption{\textbf{Overview of the proposed method}. The proposed student-teacher setup consists of two teachers and a student. The motion teacher shares motion features, whereas the denoising teacher shares denoising features. To compare the respective feature differences, perceptual losses $\calL_{\text{noise}}$ and $\calL_{\text{motion}}$ are defined. The student network has two encoders and one decoder. The final estimates are compared with the ground truth using the MSE loss $\calL_{\text{MSE}}$. }
    \label{fig:network}
\end{figure*}

\subsection{Student-Teacher Learning}
If a kernel prediction network can handle clean image sequences well and a denoising network can handle static image sequences well, is there a way we can leverage their strengths to address the dynamic low-light setting? Our solution is to develop a training scheme using the concept of student-teacher learning.

\fref{fig:network} describes our method. There are three players in this training protocol: a teacher for motion (based on kernel prediction), a teacher for denoising (based on image denoiser networks), and a student which is the network we are going to use eventually. The two teachers are individually pretrained using their respective imaging conditions. For example, the motion teacher is trained using sequences of clean and dynamic contents, whereas the denoising teacher is trained using sequences of noisy but static contents. During the training step, the teachers will transfer their knowledge to the student. During testing, only the student is used.

To transfer knowledge from the two teachers to the student, the student is first designed to have two branches, one branch duplicating the architecture of the motion teacher and another branch duplicating the architecture of the denoising teacher. When training the student, we generate three versions of the training samples. The motion teacher sees training samples that are clean and only contain motion, $\vx_{\text{motion}}$. The denoising teacher sees a training sample containing no motion but corrupted by noise, $\vx_{\text{noise}}$. The student sees the noisy dynamic sequence $\vx_{\text{QIS}}$.

Because the student has identical branches to the teachers, we can compare the features extracted by the teachers and the student. Specifically, if we denote $\phi(\cdot)$ as the feature extraction performed by the motion teacher, $\widehat{\phi}(\cdot)$ the student motion branch, $\varphi(\cdot)$ the denoising teacher, and $\widehat{\varphi}(\cdot)$ the student denoising branch, then we can define a pair of \emph{perceptual similarities}: the motion similarity
\begin{align}
\calL_{\text{motion}} = \| \underset{\text{motion student}}{\underbrace{\widehat{\phi}(\vx_{\text{QIS}})}} - \underset{\text{motion teacher}}{\underbrace{\phi(\vx_{\text{motion}})}}\|^2
\end{align}
and the denoising similarity
\begin{equation}
\calL_{\text{noise}} = \| \underset{\text{denoising student}}{\underbrace{\widehat{\varphi}(\vx_{\text{QIS}})}} - \underset{\text{denoising teacher}}{\underbrace{\varphi(\vx_{\text{noise}})}}\|^2.
\end{equation}
Intuitively, what this pair of equations does is ensure that the features extracted by the student branches are similar to those extracted by the respective teachers, which are features that can be extracted in good conditions. If this can be achieved, then we will have a good representation of the noisy dynamic sample and hence we can do a better reconstruction.

The two student branches can be considered as two autoencoders which convert the input images to codewords. As shown on the right side of \fref{fig:network}, we have a ``decoder'' which translates the concatenated codewords back to an image. The loss function of the decoder is given by the standard mean squared error (MSE) loss:
\begin{equation}
    \calL_{\text{MSE}} = \| f(\vx_{\text{QIS}}) - \vx_{\text{true}}\|^2,
\end{equation}
where $f$ is the student network and so $f(\vx_{\text{QIS}})$ denotes the estimated image. The overall loss function is the sum of these losses:
\begin{equation}
    \calL_{\text{overall}} = \calL_{\text{MSE}} + \lambda_1 \calL_{\text{motion}} + \lambda_2 \calL_{\text{noise}} ,
\end{equation}
where $\lambda_1$ and $\lambda_2$ are tunable parameters. Training the network is equivalent to finding the encoders $\widehat{\phi}$ and $\widehat{\varphi}$, and the decoder $f$.

\subsection{Choice of Teacher and Student Networks}
The proposed student-teacher framework is quite general. Specific to this paper, the two teachers and the student are chosen as follows.

The motion teacher is the kernel prediction network (KPN) \cite{mildenhall2018burst}. We modify it by removing the skip connections to maintain the information kept by the encoder. In addition, we remove the pooling layers and the bilinear upsampling layers to maximize the amount of information being fed to the feature layer. With these changes, the KPN becomes a fully convolutional-deconvolutional network.

The denoising teacher we use is a modified version of REDNet \cite{mao2016image}, which is also used in another QIS reconstruction method \cite{choi2018image}. To differentiate this single-frame REDNet and another modified version (to be discussed in the experiment section), we refer to this single-frame REDNet denoising teacher as sRED. Like the motion teacher, we remove the residual connections since they have a negative impact on the feature transfer in student-teacher learning.

The student network has two encoders and a decoder. The encoders have exactly the same architectures as the teachers. The decoder is a stack of 15 layers where each layer is a 128-channel up-convolution. The entrance layer is used to concatenate the motion and denoising features.

\section{Experiments}
\label{sec: exp}
\subsection{Experiment Settings}\label{sec: exp_set}

\textbf{Training Data}. The training data consists of two parts. The first part is for \emph{global motion}. We use the Pascal VOC 2008 dataset \cite{pascal-voc-2008} which contains 2000 training images. The second part is for \emph{local motion}. We use the Stanford Background Dataset \cite{gould2009decomposing} which contains 715 images with segmentation. For both datasets, we randomly crop patches of size $64 \times 64$ from the images to serve as ground truth. An additional 500 images are used for validation. To create global motion, we shift the patches according to a random continuous camera motion where the number of pixels traveled by the camera range from 7 to 35 across 8 consecutive frames. This is approximately 1 m/s. For local motion, we fix the background and shift the foreground using translations and rotations. The implementation of the translation is the same as that of the global motion but applied to foreground objects. The rotation is implemented by rotating the object with an angle ranging from 0 to 15 degrees.

\textbf{Training the Teachers}. The motion teacher is trained using a set of noise-free and dynamic sequences. The loss function is the mean squared error (MSE) loss suggested by \cite{mildenhall2018burst}. The network is trained for 200 epochs using the dataset described above. The denoising teacher is trained using a set of noisy but static images. Therefore, for every ground-truth sequence we generate a triplet of sequences: A noise-free dynamic sequence for the motion teacher, a noisy static image for the denoising teacher, and a noisy dynamic sequence for the student. We remark that such a data synthesis approach works for our problem because the simulated QIS data matches the statistics of real measurements.

\textbf{Baselines}. We compare the proposed methods with three existing dynamic scene reconstruction methods: (i) BM4D \cite{maggioni2012nonlocal}, (ii) Kernel Prediction Network (KPN) \cite{mildenhall2018burst}, and (iii) a modified version of REDNet \cite{mao2016image}. Our modification generalizes REDNet to multi-frame inputs, by introducing a 3D convolution at the input layer to pool the features. We refer to the modified version as multi-frame RED (mRED). Note that mRED has residual connections while sRED (denoising teacher) does not. We consider mRED a more fair baseline since it takes an input of 8 consecutive frames rather than a single frame. For KPN, the original method \cite{mildenhall2018burst} suggested using a fixed kernel size of $K=5$; we modify the setting by defining $K$ as the maximum number of pixels traveled by the motion.

\textbf{Implementation}. All networks are implemented using Keras~\cite{chollet2015keras} and TensorFlow~\cite{abadi2016tensorflow}. The student-teacher training is done using a semi-annealing process. Specifically, the regularization parameters $\lambda_1$ and $\lambda_2$ are updated once every 25 epochs such that $\lambda_1$ and $\lambda_2$ decay exponentially for the first 100 epochs. For the next 100 epochs, $\lambda_1$ and $\lambda_2$ are set to 0 and the overall loss function becomes $\calL_{\text{overall}} = \calL_{\text{MSE}}$.

\begin{figure*}[!t]
    \includegraphics[width=\linewidth]{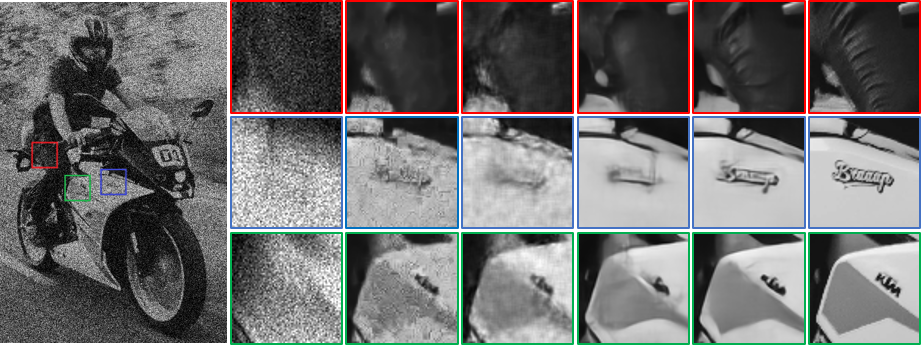} \\
    \hspace{0.5cm} \small{(a) QIS raw data, 1 frame}
    \hspace{1.2cm} \small{(b) Avg of}
    \hspace{0.75cm} \small{(c) BM4D}
    \hspace{0.75cm} \small{(d) KPN}
    \hspace{0.9cm} \small{(e) mRED}
    \hspace{0.8cm} \small{(f) Ours}
    \hspace{0.8cm} \small{(g) Ground} \\
    \hspace*{5cm} 8 frames
    \hspace{0.9cm} 23.04 dB
    \hspace{0.9cm} 25.45 dB
    \hspace{0.9cm} 26.42 dB
    \hspace{0.8cm} 29.39 dB
    \hspace{1.3cm} Truth
    \caption{\textbf{Simulated QIS data with linear global motion}. (a) The raw QIS image is simulated at 2 ppp, with a global motion of 28 pixels uniformly spaced across 8 frames. (b) An average 8 QIS raw frames. (c) BM4D \cite{maggioni2012nonlocal} (d) KPN \cite{mildenhall2018burst}. (e)  mRED, a modification of REDNet \cite{mao2016image}. (f) Proposed method. (g) Ground truth.}
    \label{fig:synthetic_results}
\end{figure*}

\begin{figure*}[!t]
\centering
\begin{tabular}{cccc}
    \hspace{-2ex}\includegraphics[width=0.23\linewidth]{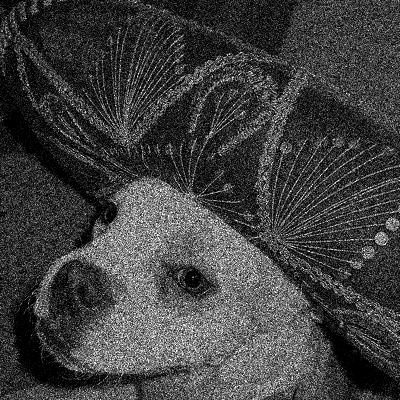}&
    \hspace{-2ex}\includegraphics[width=0.23\linewidth]{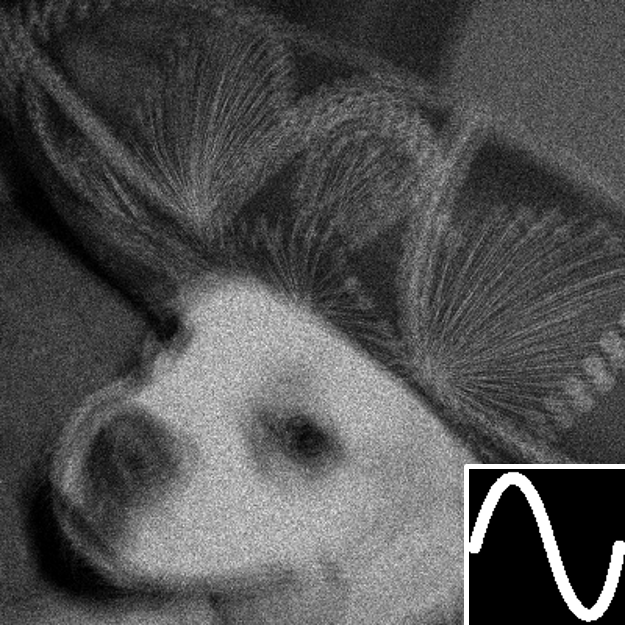}&
    \hspace{-2ex}\includegraphics[width=0.23\linewidth]{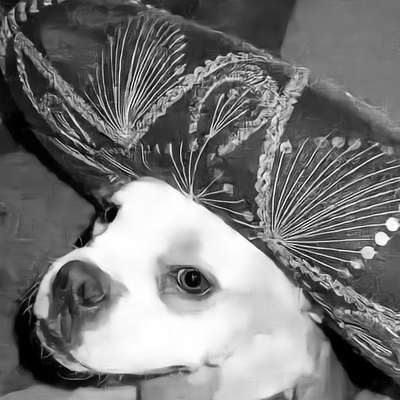}&
    \hspace{-2ex}\includegraphics[width=0.23\linewidth]{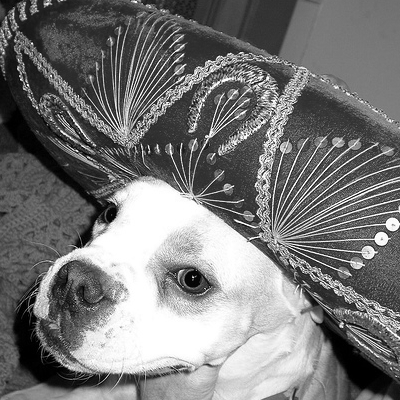}\\
    \hspace{-2ex}\includegraphics[width=0.23\linewidth]{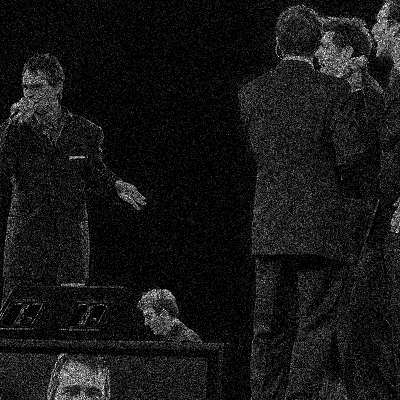}&
    \hspace{-2ex}\includegraphics[width=0.23\linewidth]{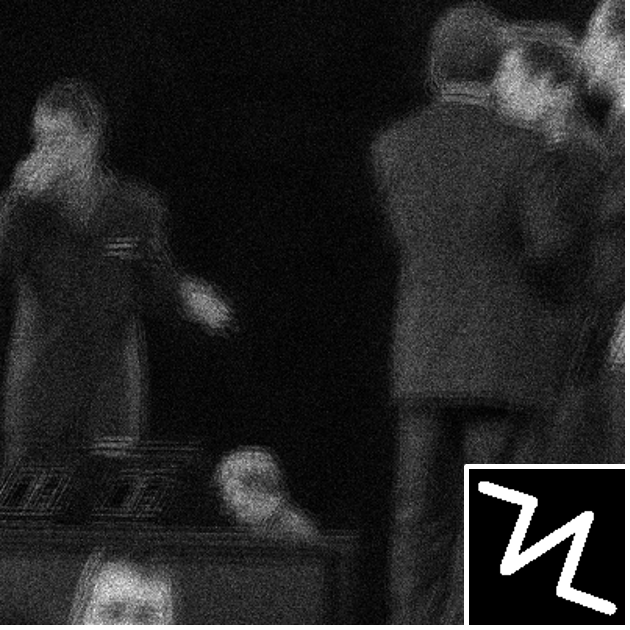}&
    \hspace{-2ex}\includegraphics[width=0.23\linewidth]{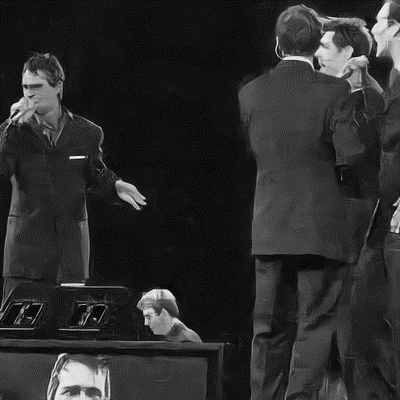}&
    \hspace{-2ex}\includegraphics[width=0.23\linewidth]{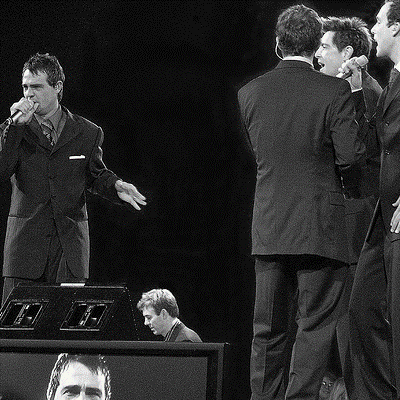}\\
    \hspace{-2ex}\includegraphics[width=0.23\linewidth]{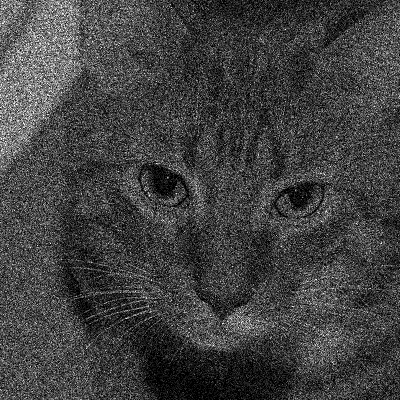}&
    \hspace{-2ex}\includegraphics[width=0.23\linewidth]{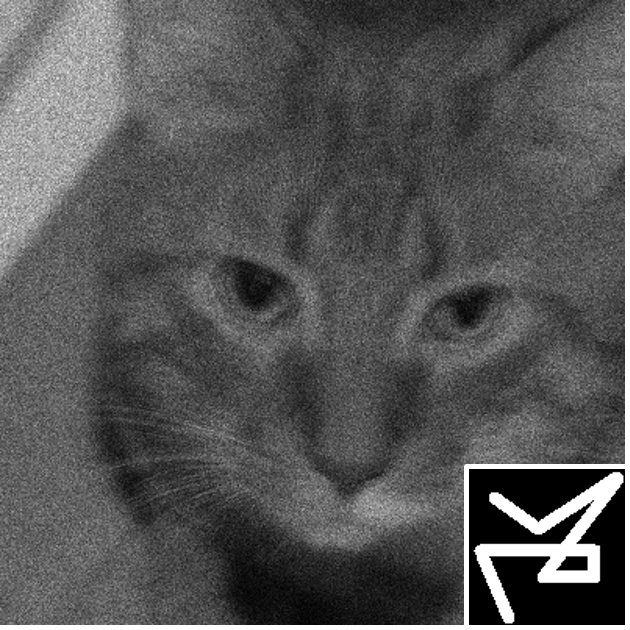}&
    \hspace{-2ex}\includegraphics[width=0.23\linewidth]{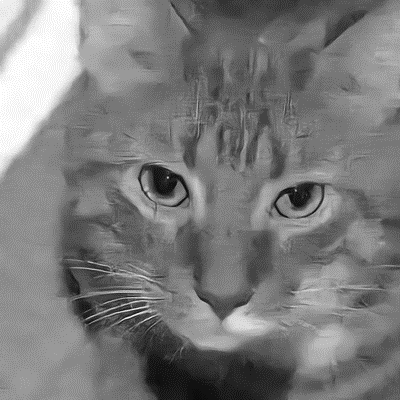}&
    \hspace{-2ex}\includegraphics[width=0.23\linewidth]{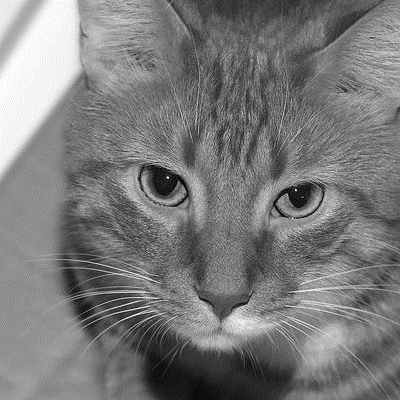}\\
    (a) QIS raw & (b) avg 8 frames & (c) Ours & (d) Ground truth
\end{tabular}
\caption{\textbf{Simulated QIS data with arbitrary global motion}. (a) QIS raw data simulated at 4 ppp. The motion trajectory is shown in the insect. (b) Average of 8 frames. (c) Proposed method. (d) Ground truth.}
\label{fig: synthetic arbitrary}
\end{figure*}

\begin{figure*}[t]
    \centering
    \begin{tabular}{cccc}
         \hspace{-2ex}\includegraphics[width=0.24\linewidth]{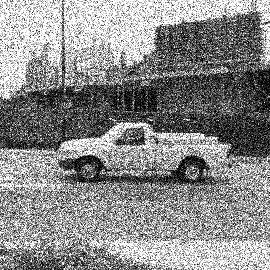} &
         \hspace{-2ex}\includegraphics[width=0.24\linewidth]{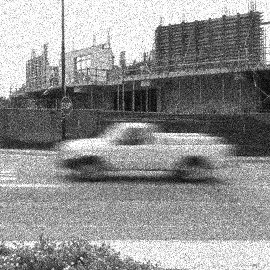} &
         \hspace{-2ex}\includegraphics[width=0.24\linewidth]{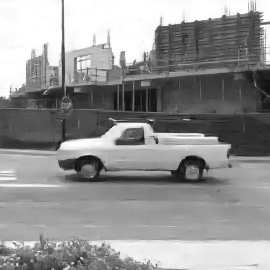} &
         \hspace{-2ex}\includegraphics[width=0.24\linewidth]{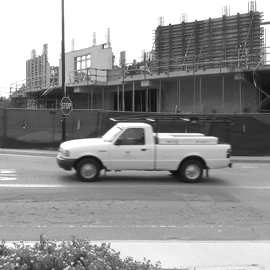} \\
         (a) QIS raw & (b) avg 8 frames & (c) Ours & (d) Ground truth
    \end{tabular}
    \caption{\textbf{Simulated QIS data with local motion}. In this example, only the car moves. The background is static. (a) Raw QIS frame assuming 1.5 ppp. (b) The average of 8 QIS frames. (c) Proposed algorithm. (d) Ground truth.}
    \label{fig:local_motion}
\end{figure*}

\begin{figure*}[t]
    \centering
    \begin{tabular}{cc}
    \hspace{-2ex}\includegraphics[trim=0cm 0cm 0.5cm 1cm, clip, width=0.5\linewidth]{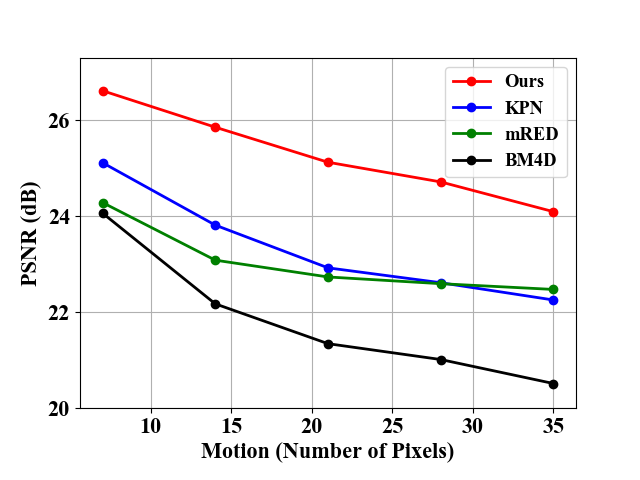} &
    \hspace{-2ex}\includegraphics[trim=0cm 0cm 0.5cm 1cm, clip, width=0.5\linewidth]{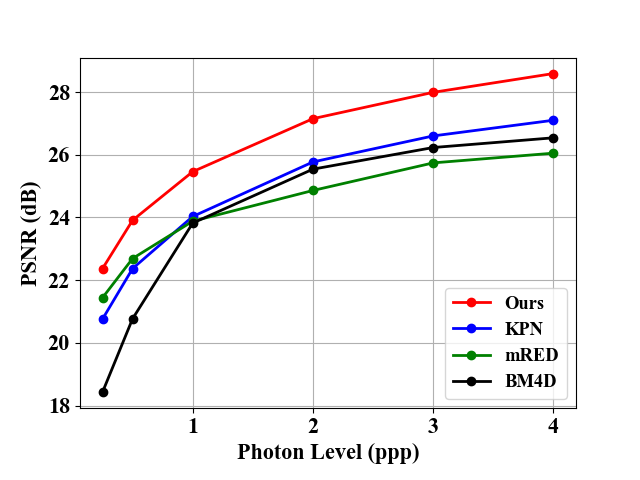}\\
    \small{(a) PSNR vs. Motion} & \small{(b) PSNR vs. Photon Level}  \\
    \small{at photon level of 2 ppp} & \small{at motion magnitude of 4 pixels}
    \end{tabular}
    \caption{\textbf{Quantitative analysis using synthetic data}. (a) PSNR as a function of the motion magnitude, at a photon level of 2 ppp. The magnitude of the motion is defined as the number of pixels traveled along the dominant direction, over 8 consecutive frames. (b) PSNR as a function of photon level. The motion magnitude is fixed at 4 pixels, but the photon level changes. Notice the consistent performance improvement of our method compared to BM4D \cite{maggioni2012nonlocal}, KPN \cite{mildenhall2018burst} and mRED (a modified version of \cite{mao2016image}).}
    \label{fig:synthetic_over}
\end{figure*}

\subsection{Synthetic Experiments}
We begin by conducting synthetic experiments. We first visually compare the reconstructed images of the proposed method and the competing methods. \fref{fig:synthetic_results} shows some results using global translation. The motion magnitude is 28 pixels across 8 frames, at 2 ppp. \fref{fig: synthetic arbitrary} shows some results using arbitrary global motion, at 4 ppp. The motion trajectory is shown in the inset in the figure. \fref{fig:local_motion} shows some results of local motion. We simulate QIS data with a real motion video of 30 fps. The photon level is 1.5 ppp. The average inference time of KPN on a $512 \times 512$ patch is 0.0886 seconds using an NVIDIA GeForce RTX 2080 Ti graphics card. For the same testing setting, mRED takes 0.0653 seconds, and the proposed method takes 0.1943 seconds. The average time for BM4D (MATLAB version) is 23.6985 seconds.

To quantitatively analyze the performance, we use the linear global motion to plot two sets of curves as shown in \fref{fig:synthetic_over}. In the first plot, we show PSNR as a function of the motion magnitude. The magnitude of the motion is defined as the number of pixels traveled along the dominant direction, over 8 consecutive frames. As shown in \fref{fig:synthetic_over}(a), the proposed method has a consistently higher PSNR compared to the three competing methods, ranging from 1.5 dB to 3 dB. This suggests that the presence of both teachers has provided a positive impact on solving the motion and noise dilemma, which is difficult for both KPN and mRED. The second set of curves is shown in \fref{fig:synthetic_over}(b) and reports PSNR as a function of the photon level. The curves in \fref{fig:synthetic_over}(b) suggest that for the photon levels we have tested, the performance gap between the proposed method and the competing methods is consistent. This provides additional evidence of the effectiveness of the proposed method.

\begin{figure*}[!t]
    \centering
    \begin{tabular}{cc}
         \hspace{-2ex}\includegraphics[width=0.47\linewidth]{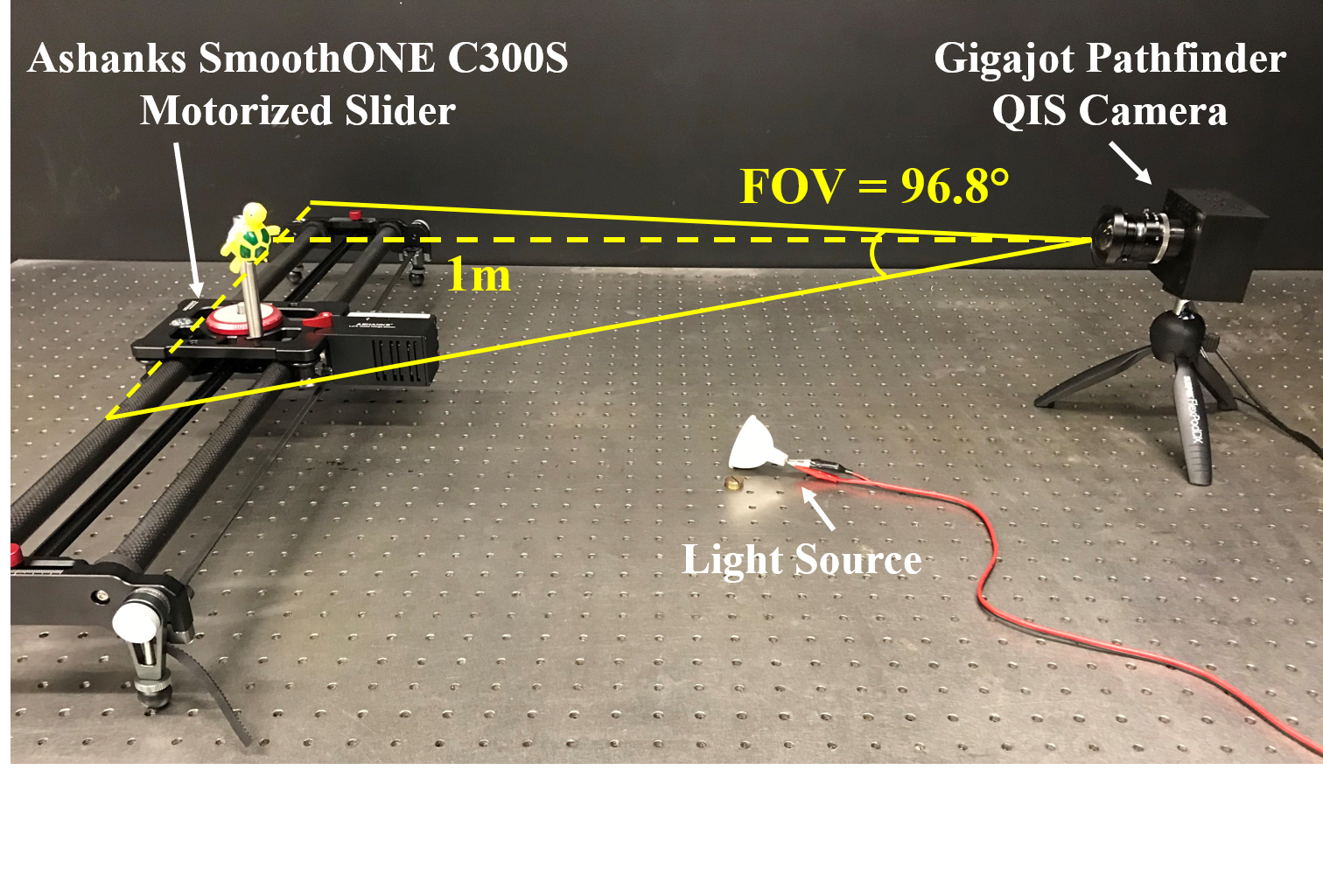} &
         \hspace{-2ex}\includegraphics[width=0.47\linewidth]{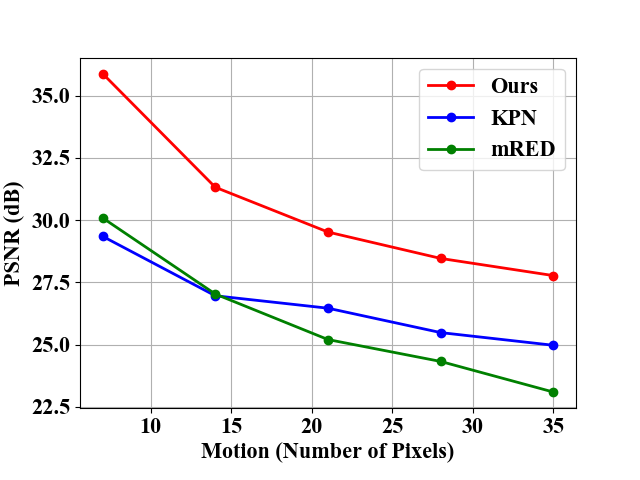}\\
         (a) Experimental Setup
         & (b) PSNR vs. motion (pixels)
    \end{tabular}
    \vspace{-1ex}
    \caption{\textbf{(a) Setup of QIS data collection}. The QIS camera is placed 1 meter from the object which is attached to a motorized slider. The horizontal field of view (FOV) of the lens is $96.8\degree$. The motion is continuous but slow. \textbf{(b) Quantitative analysis on real data}. The plot shows the PSNR values as a function of the motion magnitude, under a photon level of 0.5 ppp. The ``reference'' in this experiment is determined by reconstructing an image using a stack of static frames of the same scene. The reconstruction method is based on \cite{choi2018image}.}
    \label{fig:lab}
\end{figure*}

\begin{figure*}[!t]
    \centering
    \begin{tabular}{cccccc}
         \hspace{-2ex}\includegraphics[width=0.15\linewidth]{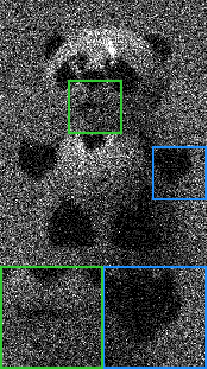} &
         \hspace{-2ex}\includegraphics[width=0.15\linewidth]{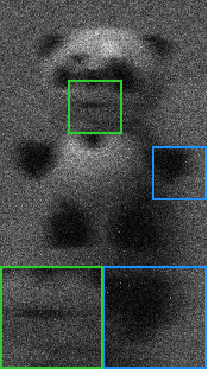} &
         \hspace{-2ex}\includegraphics[width=0.15\linewidth]{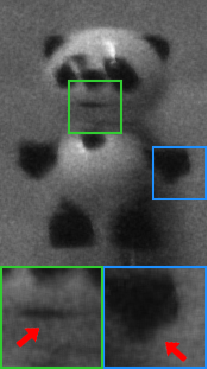} &
         \hspace{-2ex}\includegraphics[width=0.15\linewidth]{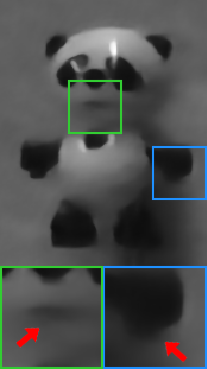} &
         \hspace{-2ex}\includegraphics[width=0.15\linewidth]{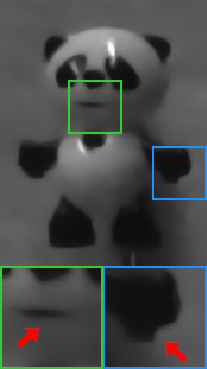} &
         \hspace{-2ex}\includegraphics[width=0.15\linewidth]{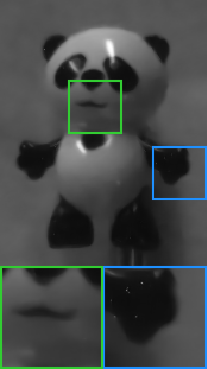} \\
         \hspace{-2ex}\includegraphics[width=0.15\linewidth]{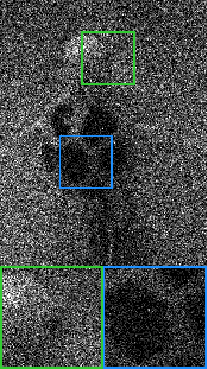} &
         \hspace{-2ex}\includegraphics[width=0.15\linewidth]{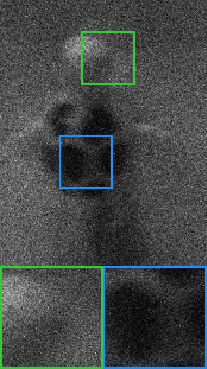} &
         \hspace{-2ex}\includegraphics[width=0.15\linewidth]{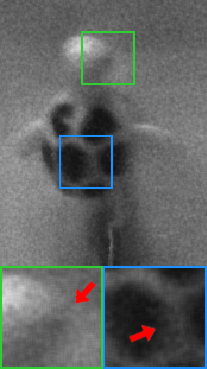} &
         \hspace{-2ex}\includegraphics[width=0.15\linewidth]{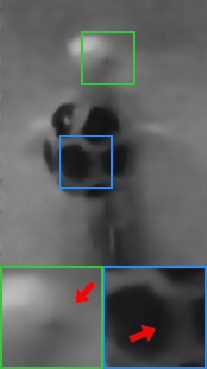} &
         \hspace{-2ex}\includegraphics[width=0.15\linewidth]{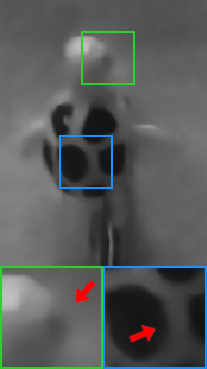} &
         \hspace{-2ex}\includegraphics[width=0.15\linewidth]{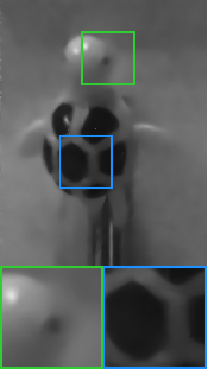} \\
         (a) QIS raw & (b) Average & (c) KPN & (d) mRED & (e) Ours & (f) Reference \\
         1 frame & 8 frames & 25.08 dB & 25.33 dB & 30.97 dB &
    \end{tabular}
    \caption{\textbf{Real QIS data}. (a) A snapshot of a real QIS frame captured at 2 ppp per frame. The number of pixels traveled by the object over the 8 frames is 28 pixels. (b) The average of 8 QIS frames. Notice the blur in the image. (c) Reconstruction result of KPN \cite{mildenhall2018burst}. (d) Reconstruction result of mRED, a modification of \cite{mao2016image}. (e) Our proposed method. (f) Reference image is a static scene denoised using mRED.}
    \label{fig:real_results}
\end{figure*}

\subsection{Real Experiments}
We verify the results using real QIS data. The real data is collected using a prototype Gigajot PathFinder camera \cite{ma2017photon}. The camera has a spatial resolution of $1024 \times 1024$. The integration time of each frame is 75 $\mu$s. Each reconstruction is based on 8 consecutive QIS frames. At the time this experiment is conducted, the readout circuit of this camera is still a prototype that is not optimized for speed. Thus, instead of demonstrating a real high-speed video, we capture a slowly moving real dynamic scene where the motion is continuous but slow. We make the exposure period short so that it is equivalent to a high-speed video. We expect that the problem will be solved in the next generation of QIS.

\begin{figure*}[t]
    \centering
    \begin{tabular}{ccc}
    \includegraphics[width=0.3\linewidth]{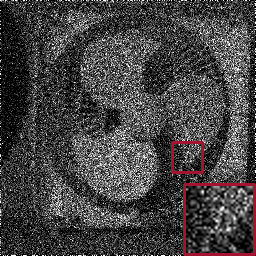} &
    \includegraphics[width=0.3\linewidth]{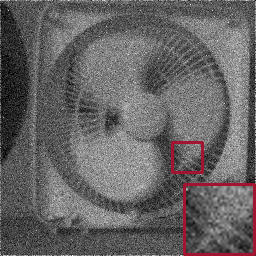} &
    \includegraphics[width=0.3\linewidth]{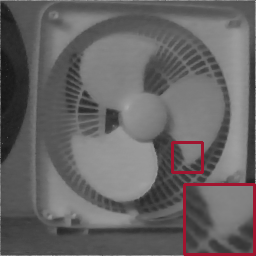} \\
    \small{(a) Real image by QIS} & \small{(b) Real image by QIS} & \small{(c) Our reconstruction} \\
    \small{1 frame, 1.5 ppp} & \small{avg of 8 frames, 1.5 ppp} & \small{using 8 QIS frames}
    \end{tabular}
    \caption{\textbf{Real QIS data with rotational motion}. The image is captured at 1.5 ppp. Notice the rotation blur in the 8-frame average, and the reconstructed result. }
    \label{fig:fan}
\end{figure*}

\begin{figure*}[t]
    \centering
    \begin{tabular}{ccccc}
        \hspace{-2ex}\includegraphics[width=0.19\linewidth]{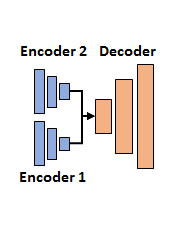}&
        \hspace{-2ex}\includegraphics[width=0.19\linewidth]{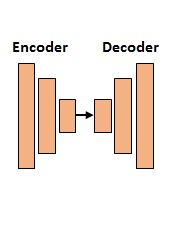}&
        \hspace{-2ex}\includegraphics[width=0.19\linewidth]{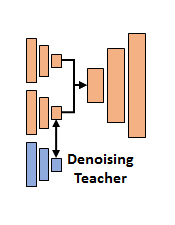}&
        \hspace{-2ex}\includegraphics[width=0.19\linewidth]{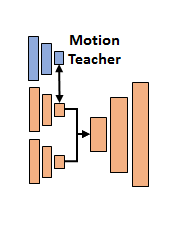}&
        \hspace{-2ex}\includegraphics[width=0.19\linewidth]{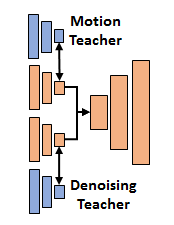}\\
        (a) Config A & (b) Config B & (c) Ours-I & (d) Ours-II & (e) Ours-full
    \end{tabular}
    \caption{\textbf{Configurations for ablation study}. (a) Config-A: Uses pre-trained teachers. (b) Config-B: Uses a single encoder instead of two smaller encoders. (c) Ours-I: Uses denoising teacher only. (d) Ours-II: Uses motion teacher only. (e) Our-full: The complete model. In this figure, blue colored layers are pre-traiend and fixed. Orange layers are trainable. }
    \label{fig:ablation}
\end{figure*}

The physical setup of the experiment is shown in \fref{fig:lab}(a). We put the camera approximately 1 meter away from the objects. The photon level is controlled by a light source. To create motion, the objects are mounted on an Ashanks SmoothONE C300S motorized camera slider, which allows us to control the location of the objects remotely. The ``ground truth'' (reference images) in this experiment is obtained by capturing a static scene via 8 consecutive QIS frames. Since these static images are noisy (due to photon shot noise), we apply mRED to denoise the images before using them as the references.

A visual comparison for this experiment is shown in \fref{fig:real_results}. The quantitative analysis is shown in \fref{fig:lab}(b), where we plot the PSNR curves as functions of the number of pixels traveled by the object. As we can see, the performance of the proposed method and the competing methods are similar to those reported in the synthetic experiments. The gap appears to be consistent with the synthetic experiments. An additional real data experiment is shown in \fref{fig:fan}, where we use QIS to capture a rotating fan scene.

\vspace{-2ex}
\subsection{Ablation Study}

We conduct an ablation study to evaluate the significance of the proposed student-teacher training protocol. \fref{fig:ablation} summarizes the 5 configurations we study. Config A is a vanilla baseline where the denoising and motion teachers are pretrained. Config B uses a single encoder instead of two encoders. Ours-I uses a student-teacher setup to train the denoising encoder. Ours-II is similar to Ours-I, but we use the motion teacher in lieu of the denoising teacher. Ours-full uses both teachers. All networks are trained using the same set of noisy and dynamic sequences. The experiments are conducted using synthetic data, at a photon level of 1 ppp and motion of 28 pixels across 8 frames. The results are summarized in \tref{tab:ablation}.

\textbf{Is student-teacher training necessary?} Configurations A and B do not use any teacher. Comparing with Ours-full, the PSNR values of Config A and Config B are worse by more than 1dB. Even if we compare with a single teacher, e.g., Ours-I, it is still 0.8dB ahead of Config B. Therefore, the student-teacher training protocol has a positive impact on performance.

\textbf{Do teacher encoders extract meaningful information?} Config A uses two pretrained encoders and a trainable decoder. The network achieves 21.51dB, which means that some features are useful for reconstruction. However, when comparing with Ours-full, it is substantially worse (23.87dB compared to 21.51dB). Since the network architectures are identical, the performance gap is likely caused by the training protocol. This indicates that the student-teacher setup is a better way to transfer knowledge from teachers to a student network.

\textbf{Which teacher to use?} Ours-I and Ours-II both use one teacher. The results suggest that if we only use one teacher, the motion teacher has a small gain (0.1dB) over the denoising teacher. However, if we use both teachers as in the proposed method, we observe another 0.2dB improvement. Thus, the presence of both teachers is helpful.

\begin{table*}[t]
    \centering
    \begin{tabular}{cccc}
        \hline
        \hline
        Configuration & \hspace{2ex} \# of Encoders \hspace{2ex} & \hspace{2ex} Which Teacher? \hspace{2ex} & Test PSNR \\
        \hline
        A & 2 & None & 21.51 dB \\
        B & 1 & None & 22.74 dB \\
        Ours-I & 2 & Denoising & 23.53 dB\\
        Ours-II & 2& Motion    & 23.65 dB\\
        Ours-full & 2 & Both   & 23.87 dB\\
        \hline
    \end{tabular}
    \vspace{1.0ex}
    \caption{\textbf{Ablation Study Results}. This table summarizes the influence of different teachers on the proposed method. The experiments are conducted using synthetic data, at a photon level of 1 ppp and a motion of 28 pixels along the dominant direction.}
    \label{tab:ablation}
    \vspace{-4ex}
\end{table*}

\section{Conclusion}
Dynamic low-light imaging is an important capability in application such as autonomous driving, security, and health care. CMOS image sensors (CIS) have fundamental limitations due to their inability to count photons. This paper considers Quanta Image Sensors (QIS) as an alternative solution. By developing a deep neural network using a new student-teacher training protocol, we demonstrated the effectiveness of transferring knowledge from a motion teacher and a denoising teacher to the student network. Experimental results indicate that the proposed method outperforms existing solutions trained under the same conditions. The proposed student-teacher protocol can also be applied to CIS problems. However, at a photon level of 1 photon per pixel or lower, QIS are necessary. Future work will focus on generalizing the reconstruction to more complex motions.

\bibliographystyle{IEEEbib}
\bibliography{egbib}

\end{document}

%% file: TeX/commands.tex
\usepackage{setspace,cite}
\usepackage{multirow,multicol}
\usepackage[tbtags]{amsmath}
\usepackage{amsbsy}
\usepackage{amssymb}
\usepackage{amsfonts}

{\begin{list}               
    {$\bullet$ \hfill}{
        \setlength{\leftmargin}{\parindent}
        \setlength{\parsep}{0.04\baselineskip}
        \setlength{\itemsep}{0.5\parsep}
        \setlength{\labelwidth}{\leftmargin}
        \setlength{\labelsep}{0em}}
    }
{\end{list}}

\providecommand{\eref}[1]{\eqref{#1}}  
\providecommand{\cref}[1]{Chapter~\ref{#1}}

\providecommand{\fref}[1]{Figure~\ref{#1}}
\providecommand{\tref}[1]{Table~\ref{#1}}

\renewcommand{\vec}[1]{\ensuremath{\boldsymbol{#1}}}

\providecommand{\calL}{\mathcal{L}}

\providecommand{\vx}{\vec{x}}

\providecommand{\veta}{\vec{\eta}}

